\documentclass[twocolumn,10pt,amsmath,amssymb,prl]{revtex4-1}
\usepackage{textcomp}
\usepackage{graphicx}
\usepackage{longtable}
\usepackage{color}
\usepackage{bm}% bold math
\usepackage{dcolumn}
\usepackage{epstopdf}
\usepackage{units}

\begin{document}

\title{Proposal for laser-cooling of rare-earth ions}

\author{Maxence Lepers$^{1}$, Ye Hong$^{1}$, Jean-Fran{\c c}ois Wyart$^{1,2}$ and Olivier Dulieu$^{1}$}
\affiliation{${}^{1}$Laboratoire Aim\'e Cotton, CNRS/Universit\'e Paris-Sud/ENS-Cachan, B\^at.~505, Campus d'Orsay, 91405 Orsay, France}
\email{maxence.lepers@u-psud.fr}
\affiliation{${}^{2}$LERMA, UMR8112, Observatoire de Paris-Meudon, Universit\'e Pierre et Marie Curie, 92195 Meudon, France}

\date{\today}

\begin{abstract}
The efficiency of laser-cooling relies on the existence of an almost closed optical-transition cycle in the energy spectrum of the considered species. In this respect rare-earth elements exhibit many transitions which are likely to induce noticeable leaks from the cooling cycle. In this work, to determine whether laser-cooling of singly-ionized erbium Er$^+$ is feasible, we have performed accurate electronic-structure calculations of energies and spontaneous-emission Einstein coefficients of Er$^+$, using a combination of \textit{ab initio} and least-square-fitting techniques. We identify five weak closed transitions suitable for laser-cooling, the broadest of which is in the kilohertz range. For the strongest transitions, by simulating the cascade dynamics of spontaneous emission, we show that repumping is necessary, and we discuss possible repumping schemes. We expect our detailed study on Er$^+$ to give a good insight into laser-cooling of neighboring ions like Dy$^+$. 
\end{abstract}

\maketitle

%\paragraph*{Introduction.}

Rare-earth elements are currently widespread in many areas of industry, including telecommunications and electronics. Over the last years they have also entered the field of ultracold matter, for which they present very suitable properties \cite{mcclelland2006, lu2010, sukachev2010, miao2014}. For example the strong magnetic moment of open-$4f$-shell neutral lanthanide atoms, up to 10 Bohr magnetons $\mu_\mathrm{B}$ for dysprosium, induces anisotropic and long-range dipole-dipole interactions, which make those atoms excellent candidates for the production of ultracold dipolar gases \cite{baranov2008, lahaye2009, newman2011, aikawa2014b, burdick2015, frisch2015}. In particular erbium and dysprosium present bosonic and fermionic stable isotopes, for which Bose-Einstein condensation and Fermi degeneracy were achieved \cite{lu2011b, aikawa2012, lu2012, aikawa2014}.

Meanwhile, laser-cooling of trapped ions \cite{eschner2003} has allowed for reaching an exceptional control on quantum systems, down to the single-particle level \cite{leibfried2003, wineland2013}. This led to the realization of high-precision optical clocks \cite{diddams2001, schneider2005, rosenband2008, chwalla2009, huntemann2012, dube2014, godun2014, ludlow2015}, or of logic gates for quantum-information processing \cite{kielpinski2002, gulde2003, blatt2008, home2009, monroe2013}. Another noticeable trend in cold matter is to merge cold-atom and cold-ion traps, in order to study elementary chemical reactions, like charge transfer or molecular-ion formation \cite{willitsch2008, schmid2010, zipkes2010b, hall2011, rellegert2011, da-silva2015}. Up to now all laser-cooled ions have a similar electronic structure: most often one, or possibly two valence electron around a closed-shell core, like in alkaline-earth, mercury (Hg$^+$), ytterbium (Yb$^+$) or indium (In$^+$) \cite{peik1994}. Laser-cooling of solids doped with rare-earth ions was also reported in several experiments \cite{nemova2010}.

\begin{figure}
\begin{centering}
\includegraphics[width=8.5cm]{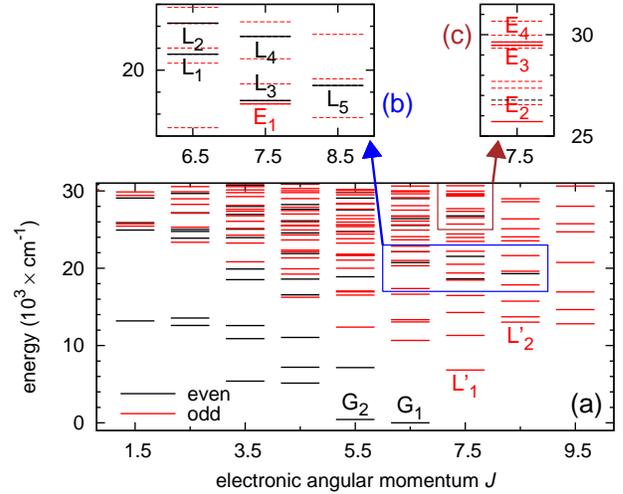}
\par\end{centering}
\caption{(Color online) (a) Experimentally observed \cite{wyart2009} energy levels of Er$^+$ sorted by parity, as functions of the total electronic angular momentum $J$. In panels (b) and (c), which are zooms on two areas of (a), the levels in solid lines are relevant for laser-cooling, while those in dashed lines are not.
\label{fig:ener-lev}}
\end{figure}

In this Letter we propose a scheme for laser-cooling of open-$4f$-shell rare-earth ions, taking the example of singly-ionized erbium Er$^+$, whose rich electronic structure yields advantageous properties for ultracold-matter physics. For example Er$^+$ spectrum is characterized by a forest of weak radiative transitions from which emerge a few strong transitions \cite{xu2003, stockett2007, lawler2008}, adapted to laser-cooling and trapping. Moreover the first excited level of Er$^+$, denoted G$_2$, lies only 440 cm$^{-1}$ above the ground level G$_1$ (see Fig.~\ref{fig:ener-lev} and Table \ref{tab:ener-lev}); and the G$_1$-G$_2$ transition is allowed both in the electric-quadrupole (E2) and magnetic-dipole (M1) approximations, with widths equal to 0.1 mHz and 0.9 pHz respectively (see below). The electronic structure of the levels G$_1$ and G$_2$ is so similar, that their static dipole polarizabilities only differ by at most 0.2 \% (see below), which makes the G$_1$-G$_2$ transition weakly sensitive to differential Stark shifts \cite{kozlov2013}. Moreover the strong magnetic moment of Er$^+$ equal to 8~$\mu_\mathrm{B}$, opens the possibility to observe magnetic-dipole interactions \cite{kotler2014}.

Nevertheless the rich structure of Er$^+$ makes it \textit{a priori} difficult to find closed cycles of strong transitions. Excited ions can decay towards many leaking levels, which gives rise to a cascade dynamics that potentially recycles the ions to the ground level. In order to determine the feasibility of Er$^+$ laser-cooling, we have thus made accurate quantum-chemical calculations of Einstein-$A$ coefficients, characterizing spontaneous emission from excited level suitable for laser-cooling towards the ground level, but also to all possible leaking levels. By inserting the computed $A$ coefficients in a model giving the time-dependent fraction of ions in each electronic state, we show that the recycling mechanism found in neutral erbium \cite{mcclelland2006} is not efficient enough here, and so we identify possible repumping schemes based on at least two auxiliary transitions.

\begin{table}
\caption{Characteristics of the energy levels $i$ relevant for laser-cooling (see Fig.~\ref{fig:ener-lev} and Ref.~\cite{wyart2009}): total electronic angular momentum $J_i$, leading electronic configuration (where the xenon core [Xe] has been omitted), leading multiplet, experimental and theoretical energies $E_i^\mathrm{exp}$ and $E_i^\mathrm{th}$. The multiplets of levels G$_1$, G$_2$, L$'_1$, L$_1$--L$_5$ and E$_4$ come from the NIST database \cite{NIST-ASD}, while for the other levels, they are extracted from our eigenvector calculations.
\label{tab:ener-lev}}
\begin{ruledtabular}
\begin{tabular}{crccrr}
   label & $J_i$ & leading        & leading & $E_i^\mathrm{exp}$
                                         & $E_i^\mathrm{th}$ \\
    $i$  &       & configuration  & multiplet & (cm$^{-1}$) 
                                         & (cm$^{-1}$) \\
  \hline
   G$_1$  & $13/2$ 
      & $4f^{12}(^3H_6)6s(^2S_{1/2})$ & $(6,\nicefrac{1}{2})$
      &     0 &    14 \\
   G$_2$  & $11/2$ 
      & $4f^{12}(^3H_6)6s(^2S_{1/2})$ & $(6,\nicefrac{1}{2})$
      &   440 &   447 \\
   L$'_1$ & $15/2$ & $4f^{11}6s^2$ & $^4I^o$     
      &  6825 &  6796 \\
   L$'_2$ & $17/2$ 
      & $4f^{11}(^4I_{15/2}^o)5d6s(^3D_1)$ & $(\nicefrac{15}{2},1)^o$ 
      & 13028 & 12948 \\
   L$_1$  & $13/2$ 
      & $4f^{12}(^3H_6)5d(^2D_{3/2})$ & $(6,\nicefrac{3}{2})$
      & 20728 & 20601 \\
   L$_2$  & $13/2$ 
      & $4f^{12}(^3H_6)5d(^2D_{5/2})$ & $(6,\nicefrac{5}{2})$
      & 22141 & 22154 \\
   L$_3$  & $15/2$ 
      & $4f^{12}(^3H_6)5d(^2D_{3/2})$ & $(6,\nicefrac{3}{2})$
      & 18617 & 18655 \\
   L$_4$  & $15/2$ 
      & $4f^{12}(^3H_6)5d(^2D_{5/2})$ & $(6,\nicefrac{5}{2})$
      & 21533 & 21601 \\
   L$_5$  & $17/2$ 
      & $4f^{12}(^3H_6)5d(^2D_{5/2})$ & $(6,\nicefrac{5}{2})$
      & 19303 & 19302 \\
   E$_1$  & $15/2$ 
      & $4f^{11}(^4I_{15/2}^o)5d6s(^1D_2)$ & $(\nicefrac{15}{2},2)^o$ 
      & 18463 & 18412 \\
   E$_2$  & $15/2$ 
      & $4f^{11}(^4I_{11/2}^o)5d6s(^3D_3)$ & $(\nicefrac{11}{2},3)^o$        
      & 25712 & 25761 \\
   E$_3$  & $15/2$ 
      & $4f^{11}(^4I_{15/2}^o)5d^2$ & $^o$ 
      & 29473 & 29491 \\
   E$_4$  & $15/2$ 
      & $4f^{12}(^3H_6)6p(^2P_{3/2}^o)$ & $(6,\nicefrac{3}{2})^o$
      & 29641 & 29655 \\
\end{tabular}
\end{ruledtabular}
\end{table}

%\paragraph*{Calculations of energies and Einstein-$A$ coefficients.}

Our electronic-structure calculations, which were performed with the Racah-Slater method implemented the Cowan codes \cite{cowan1981}, were described in Refs.~\cite{wyart2009, wyart2011, lepers2014}. They are composed of three steps. (i) Energies and Einstein coefficients are computed \textit{ab initio} using a Hartree-Fock method including relativistic corrections and combined with configuration interaction (HFR+CI). For each parity, the calculated energies depend on quantities like direct and exchange Coulombic integrals, which are characteristic of each electronic configuration. (ii) These quantities are treated as adjustable parameters in order to fit the theoretical energies to the experimental ones by the least-square method. (iii) Similarly to energies, our theoretical $A$ coefficients depend on a restricted number of radiative parameters that are also adjusted by fitting our $A$ coefficients with available experimental ones \cite{ruczkowski2014, bouazza2015}.
In the case of Er$^+$, steps (i) and (ii) resulted in the detailed interpretation of the energy spectrum given in Ref.~\cite{wyart2009}. For the evaluation of $A$ coefficients (step (iii)), the method of \cite{lepers2014}, limited up to now to the transitions involving only the ground level, is now extended to transitions involving an arbitrary number of levels in both parities.

The electronic configurations included in our model are: $\mathrm{[Xe]}4f^{12}6s$, $\mathrm{[Xe]}4f^{12}5d$, $\mathrm{[Xe]}4f^{11}6s6p$ and $\mathrm{[Xe]}4f^{11}5d6p$ for even-parity levels, and $\mathrm{[Xe]}4f^{11}6s^2$, $\mathrm{[Xe]}4f^{12}6p$, $\mathrm{[Xe]}4f^{11}5d6s$, $\mathrm{[Xe]}4f^{13}$ and $\mathrm{[Xe]}4f^{11}5d^2$ for odd-parity levels, $\mathrm{[Xe]}$ denoting the electronic configuration of xenon, omitted in what follows. For example, among the computed even-parity levels, 130 were fitted to their known experimental counterparts, using 25 free energetic parameters, giving a 62-cm$^{-1}$ standard deviation.

After steps (i) and (ii) the level $i$ is described by a CI wave function $|i\rangle=\sum_p c_{ip}|p\rangle$, where $|p\rangle$ formally represents an electronic configuration. The theoretical Einstein coefficients $A_{ij}^\mathrm{th}$ characterizing the probability of spontaneous emission from level $j$ to level $i$ can be expanded
\begin{equation}
  A_{ij}^\mathrm{th} = \left( \sum_{pq} a_{ij,pq}
    \left\langle n\ell,p\right|
    \hat{r}\left|n'\ell',q\right\rangle 
  \right)^2 ,
  \label{eq:aik}
\end{equation}
where the configurations $p$ and $q$ are identical, except for one electron that hops from subshells $n'\ell'$ to $n\ell$, with $\ell'-\ell=0,\,\pm1$ for electric-dipole (E1) transitions. Unlike the coefficients $a_{ij,pq}$ which are specific to each transition, the matrix elements of the monoelectronic $\hat{r}$-operator are common parameters to all transitions. 

The configurations included in our model give rise to ten possible $\hat{r}$ matrix elements: three for 
$(n\ell$-$n'\ell') = (6s$-$6p)$ transitions, namely $(p,q)=(4f^{12}6s,4f^{12}6p)$, $(4f^{11}5d6s,4f^{11}5d6p)$ and $(4f^{11}6s^2,4f^{11}6s6p)$, three for $6p$-$5d$, and four for $5d$-$4f$ transitions. In step (iii) we fitted the scaling factors (SFs) $f_m = \left\langle n\ell,p\right| \hat{r}\left|n'\ell',q\right\rangle / \left\langle n\ell,p\right| \hat{r}\left| n'\ell',q\right\rangle_\mathrm{HFR}$ between $\hat{r}$ matrix elements and their computed HFR values, with $m\equiv (n\ell pn'\ell'q)$, to have the best agreement between experimental and theoretical $A$ coefficients. Among the ten scaling factors, those corresponding to $6p$-$5d$ transitions on one hand, and to $5d$-$4f$ transitions on the other hand, were constrained to be equal; they are called $f_4$ and $f_5$ respectively. On the contrary, the three scaling factors corresponding to the strong $6s$-$6p$ transitions can vary independently; they are called $f_1$, $f_2$ and $f_3$ for $(p,q)=(4f^{12}6s,4f^{12}6p)$, $(4f^{11}5d6s,4f^{11}5d6p)$ and $(4f^{11}6s^2,4f^{11}6s6p)$ respectively.

We fitted the SFs $f_m$ using the experimental Einstein coefficients given by Lawler \textit{et al.}~\cite{lawler2008}.
Due to strong differences between $A_{ij}^\mathrm{th}$ and $A_{ij}^\mathrm{exp}$, we excluded 17 of the 418 transitions of Ref.~\cite{lawler2008}. We checked that none of them were involved in the laser-cooling process discussed here. To account for the uncertainty of measurements we made 100 fits in which all the $A_{ij}^\mathrm{exp}$ coefficients have a random value within their uncertainty range. Averaging the best $f_m$ obtained for each of the 100 shots, we obtain finally the optimal SFs: $f_1=0.887$, $f_2=0.798$, $f_3=0.921$, $f_4=0.840$ and $f_5=0.817$, which give a standard deviation (see \cite{lepers2014}, Eq.~(15)) $\sigma_A=4.18\times 10^6$ s$^{-1}$, with $N_\mathrm{par}=5$ free SFs, and $N_\mathrm{lev}=401$ transitions. With those optimal SFs, 40 \% of the $A$ coefficient are calculated with a precision better than 12 \%. A satisfactory comparison of experimental and theoretical $A$ coefficients is presented in Table \ref{tab:aik} for a selection of transitions relevant for laser-cooling. With this optimal set of $A$ coefficients we can calculate the scalar static dipole polarizability of any level, using Eq.~(4) of Ref.~\cite{lepers2014}, which gives, for G$_1$ and G$_2$, the same value of 59.4 $a_0^3$, $a_0$ being the Bohr radius.

In order for instance to characterize the G$_1$-G$_2$ transition, the Cowan codes also allow for calculating Einstein coefficients of E2 and M1 transitions. In principle E2 Einstein coefficients can be determined using the procedure described above, by changing $\hat{r}$ into $\hat{r}^2$ in Eq.~(\ref{eq:aik}), and taking configurations $p$ and $q$ in the same parity and $\ell-\ell'=0$, $\pm1$ or $\pm2$. But due to the absence of experimental data we cannot apply step (iii), and we calculate the E2 $A$-coefficients with HFR $\hat{r}^2$ matrix elements.

\begin{table}
\caption{Einstein-$A$ coefficients characterizing the spontaneous emission from laser-cooling levels E$_2$--E$_4$ to the ground level G$_1$ and to the main leaking levels L$_1$--L$_5$. Our theoretical values are compared with available experimental ones \cite{lawler2008}. The notation $(n)$ stands for $\times 10^n$.
\label{tab:aik}}
\begin{ruledtabular}
\begin{tabular}{c|rrrrr}
   $j$ (odd) $\to$ & \multicolumn{1}{c}{E$_2$} 
                   & \multicolumn{2}{c}{E$_3$} 
                   & \multicolumn{2}{c}{E$_4$} \\
   $\Gamma_j^\mathrm{th}$ (s$^{-1}$) & 1.058(6) 
                   & \multicolumn{2}{c}{1.533(7)} 
                   & \multicolumn{2}{c}{1.546(8)} \\
  \hline
   $i$                     & $A_{ij}^\mathrm{th }$ 
   & $A_{ij}^\mathrm{exp}$ & $A_{ij}^\mathrm{th }$ 
   & $A_{ij}^\mathrm{exp}$ & $A_{ij}^\mathrm{th }$ \\
   (even) $\downarrow$ & (s$^{-1}$) & (s$^{-1}$) & (s$^{-1}$)
                       & (s$^{-1}$) & (s$^{-1}$) \\
  \hline
   G$_1$ & 1.03(6) & 2.83(7) & 1.48(7) & 1.45(8) & 1.49(8) \\
   L$_1$ & 8.58(2) &  5.4(4) & 4.16(4) &  1.4(5) & 1.58(5) \\
   L$_2$ & 9.93(1) &       - & 6.01(3) &  6.5(4) & 5.03(4) \\
   L$_3$ & 1.65(3) &  1.9(5) & 1.53(5) &  1.9(5) & 1.57(5) \\
   L$_4$ & 1.69(3) &  8.6(4) & 3.31(4) &  8.6(4) & 6.15(5) \\
   L$_5$ & 2.61(4) &  6.6(5) & 2.46(5) &  5.3(6) & 4.25(6) \\
\end{tabular}
\end{ruledtabular}
\end{table}

%\paragraph*{Suitable laser-cooling transitions and possible leaking pathways.}

In order to avoid leaks towards the levels of the lowest configuration, including G$_2$, we consider cooling transitions with odd-parity upper levels of total angular momentum $J_i=15/2$. We identify 18 possible candidates corresponding to transition energy below 30000 cm$^{-1}$. Among them five levels give rise to closed transitions with the ground level, but with small $A$ coefficients. For instance, the energy of the upper level E$_1$ of the strongest closed transition is $E_{\mathrm{E}_1}^\mathrm{exp}=18463$ cm$^{-1}$, and the corresponding Einstein coefficient is $A_{\mathrm{G}_1 \mathrm{E}_1}^\mathrm{th} = 1.68\times 10^4$ s$^{-1}$. By contrast, the three strongest transitions, to levels E$_2$, E$_3$ and E$_4$, which respectively possess 2, 7 and 75 \% of $4f^{12}6p$ component, and which are therefore characterized by $A$ coefficients larger than $10^6$ s$^{-1}$, will be considered in details in what follows.

Regarding leaking levels, it is remarkable to note that the three levels E$_2$, E$_3$ and E$_4$ behave in a very similar way. Calculating the branching ratios (BRs) $r_{ij}=A_{ij}^\mathrm{th} /\Gamma^\mathrm{th}_j$, where $\Gamma^\mathrm{th}_j$ is the theoretical inverse lifetime of level $j$ given in Table \ref{tab:aik}, we see that levels $j=\mathrm{E}_2$, E$_3$ and E$_4$ mainly decay to the ground level $i=G_1$, with branching ratios respectively equal to 97.1 \%, 96.7 \% and 96.6 \%. Then the main source of leaks is the level L$_5$, with $r_{ij}=2.5$ \%, 1.7 \% and 2.8 \%. Levels L$_1$--L$_4$ give BRs between 0.1 and 1 \%, while levels not mentioned in Table \ref{tab:aik} give BRs below 0.001 \%. The common feature of levels L$_1$--L$_5$ is their dominant $4f^{12}(^3H_6)5d(^2D_j)\,(6,j)_{J_i}$ character, where $j=3/2$, 5/2 and $J_i=13/2$, 15/2 and 17/2, which means that the leaks are essentially due to $np_{3/2}$-$(n-1)d_{3/2,5/2}$ transitions, like in alkaline-earth ions.

%\paragraph*{Cascade dynamics.}

The difference between alkaline-earth ions and Er$^+$ is that electric-dipole transitions are possible from levels L$_1$--L$_5$ to so-called secondary  leaking levels, including L$'_1$ and L$'_2$, which can themselves decay to lower levels, and so forth. To determine the efficiency of recycling to the ground level, it is therefore necessary to trace out the time-dependent population of each level, when all lasers are off \cite{mcclelland2006}. To that end, we consider that, at initial time $t=0$, all the ions are in the excited level E$_2$, E$_3$ or E$_4$; then they cascade to lower levels by spontaneous emission, until they reach a steady state. The fraction $X_i$ of ions in level $i$ obeys the system of differential equations
\begin{equation}
  \frac{dX_i}{dt} = -\Gamma_i X_i + \sum_{j=i+1}^n A_{ij}X_j \,,
  \label{eq:xi-tde}
\end{equation}
where $\Gamma_i=\sum_{j=0}^{i-1}A_{ji}$. In Eq.~(\ref{eq:xi-tde}) the index $i$ formally denotes the $i$-th excited level of the ion. Namely, $i=0$ correspond to the ground level G$_1$, $i=1$ to the first excited level G$_2$, and $i=162$ to E$_4$. In Eq.~(\ref{eq:xi-tde}), $n$ is the initially populated level, \textit{i.e.}~$X_i(t=0)=\delta_{in}$, and at any time the ionic population is conserved \textit{i.e.}~$\sum_{i=0}^n X_i(t)=1$.
Equation (\ref{eq:xi-tde}) can be solved analytically, by starting with level $n$, which decays exponentially, $X_n=\exp(-\Gamma_nt)$, and which acts as a source for $n-1$, $n-2$, etc.. Finally we obtain the general form
\begin{equation}
  X_i(t) = \sum_{j=i}^n C_{ij}e^{-\Gamma_jt}
  \label{eq:xi-t}
\end{equation}
where the time independent coefficients $C_{ij}$ are given by the recursion relation
\begin{equation}
  C_{ij} = \frac{1}{\Gamma_i-\Gamma_j} \sum_{k=i+1}^j
    A_{ik} C_{kj}
  \label{eq:cij}
\end{equation}
defined for $i<k\le j$, and by $C_{ii}=X_i(t=0)=\delta_{in}$. 

\begin{figure}
\begin{centering}
\includegraphics[width=8.5cm]{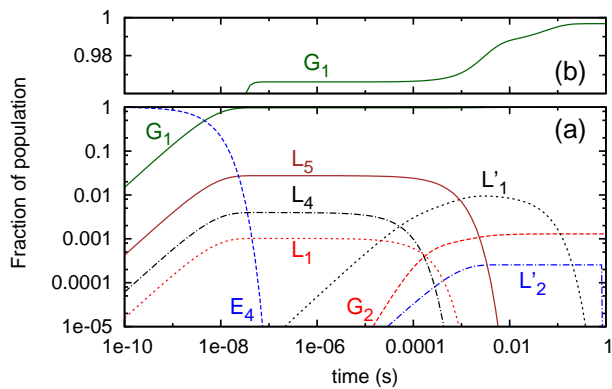}
\par\end{centering}
\caption{(Color online) (a) Time evolution of the fraction of ions in some levels relevant for laser-cooling (see Fig.~\ref{fig:ener-lev} and Table \ref{tab:ener-lev}). in log-log scale (see Eqs.~(\ref{eq:xi-t}) and (\ref{eq:cij})). Panel (b) is a zoom on fractions close to unity, presented in linear scale.
\label{fig:pop-t}}
\end{figure}

Figure \ref{fig:pop-t} shows the time evolution of the fraction of ions in some levels relevant in the cascade dynamics, for $n=162$ (level E$_4$) and considering E1 transitions. Within a fast time scale $\Gamma_{E_4}^{-1}\approx 10$ ns, the ions leave the level E$_4$ and populate G$_1$, L$_1$, L$_4$ and L$_5$. An early quasi-steady state is reached, with populations corresponding to the BRs calculated above. After a fraction a millisecond, the population of those primary leaking levels is transferred to secondary ones, some of which, including L$'_2$ and to a large extent G$_2$, definitively ``{}trap'' the ions. On the contrary other levels like L$'_1$ can empty themselves towards the ground level G$_1$ which reaches its steady state of 99.7 \% after a fraction of second (see Fig.~\ref{fig:pop-t}(b)).

\begin{table}
\caption{Percentage (\%) of ions in the three most populated levels calculated using Eqs.~(\ref{eq:xi-t}) and (\ref{eq:cij}) after $t=1$ second, only with E1 Einstein coefficients, or with E1, E2 and M1 Einstein coefficients (label {}``E/M").
\label{tab:pop-fin}}
\begin{ruledtabular}
\begin{tabular}{c|dddddd}
   & \multicolumn{2}{c}{E$_2$} 
   & \multicolumn{2}{c}{E$_3$} 
   & \multicolumn{2}{c}{E$_4$} \\
   & \multicolumn{1}{c}{$\mathrm{E1}$} 
   & \multicolumn{1}{c}{$\mathrm{E/M}$}
   & \multicolumn{1}{c}{$\mathrm{E1}$} 
   & \multicolumn{1}{c}{$\mathrm{E/M}$}
   & \multicolumn{1}{c}{$\mathrm{E1}$} 
   & \multicolumn{1}{c}{$\mathrm{E/M}$}  \\
  \hline
   G$_1$  & 99.72 & 99.73 & 99.23 & 99.25 & 99.68 & 99.71 \\
   G$_2$  &  0.12 &  0.12 &  0.62 &  0.63 &  0.12 &  0.13 \\
   L$'_2$ &  0.02 &  0.03 &  0.02 &  0.02 &  0.03 &  0.03 \\
   other  &  0.14 &  0.12 &  0.13 &  0.13 &  0.17 &  0.17 \\
\end{tabular}
\end{ruledtabular}
\end{table}

Table \ref{tab:pop-fin} gives the population fractions after 1 second, when the steady state is reached (see Fig.~\ref{fig:pop-t}), and for the three initial levels E$_2$, E$_3$ and E$_4$. In view of the slow dynamics observed on Fig.~\ref{fig:pop-t}, we have also included E2 and M1 Einstein coefficients in our rate equations (\ref{eq:xi-tde}). However their influence is not significant after 1 second, because when they could play a significant role, \textit{e.g.}~from L$_1$--L$_5$ to G$_1$ in the E2 case, E1 transitions to secondary leaking levels are actually faster. Table \ref{tab:pop-fin} also indicates that the recycling process in not sufficient to ensure an efficient cooling, since even after 1 second, a noticeable fraction of ions, 0.3 \%, is trapped in several metastable levels, including L$'_2$. The ions in L$'_2$ can decay to L$'_1$ through an E2 transition, but only after ~$10^6$ s. Therefore repumping turns out to be necessary.

%\paragraph*{Possible repumping schemes.}

Being the main source of leaks from E$_2$, E$_3$ and E$_4$, the level L$_5$ is a natural candidate for repumping, also because spontaneous-emission transitions from L$_5$ would drive the ions to odd-parity high-$J$ levels, and so would decrease the probability of recycling to the ground level. However L$_5$ cannot be the only {}``repumped'' level, since roughly 2 \% of the ions would leave the cooling cycle at each transition. One possibility would then be to use repumping from the other primary leaking levels L$_1$--L$_4$. Repumping the five levels L$_1$--L$_5$ would cause less than $10^{-5}$ of loss per transition; but using five repumping lasers seems experimentally unrealistic.

Another possibility, in addition to level L$_5$, would be to inject back to the cooling cycle the ions accumulated in level G$_2$ after a few milliseconds (see Fig.~\ref{fig:pop-t}). However direct repumping  to levels E$_2$--E$_4$ is not possible, at least below the electric-octupole approximation. A direct transfer from G$_2$ to G$_1$, \textit{e.g.}~through a $\pi$ pulse, is also doable, provided that G$_1$ is empty when the pulse is applied. The ions could also be repumped from G$_2$ to an auxiliary odd-parity level which preferentially decays to the ground level G$_1$. The level denoted E' ($J_{\mathrm{E}'}=13/2$, $E_\mathrm{E'}^\mathrm{exp}=29628$ cm$^{-1}$, $E_\mathrm{E'}^\mathrm{th}=29636$ cm$^{-1}$) seems a good candidate, since the corresponding coefficients are $A_{\mathrm{G}_1\mathrm{E'}}^\mathrm{exp}=1.52(7)$ s$^{-1}$, $A_{\mathrm{G}_1\mathrm{E'}}^\mathrm{th}=1.02(7)$ s$^{-1}$, $A_{\mathrm{G}_2\mathrm{E'}}^\mathrm{exp}=3.11(6)$ s$^{-1}$ and $A_{\mathrm{G}_2\mathrm{E'}}^\mathrm{th}=1.38(6)$ s$^{-1}$. But again a small fraction of ions excited in this level would decay to undesired leaking levels.

%\paragraph*{Conclusion.}

In this article we have addressed the feasibility of singly-ionized erbium (Er$^+$) laser-cooling, by modelling its energy spectrum and Einstein-$A$ coefficients. The most promising way is the closed transition to level E$_1$, which is much weaker than the commonly used transitions, but much stronger than the E2 transition chosen to cool Ca$^+$ in Ref.~\cite{hendricks2008}. Regarding the transitions to levels with $4f^{12}6p$ character, we observe significant leaks to levels belonging to the $4f^{12}5d$ configuration. Due to the passive role of the $4f$ electrons in the leaking process, we expect our conclusions to be valid for neighboring rare-earth ions like dysprosium, holmium or thulium.

%\paragraph*{Acknowledgments.}

M.~L. and O.~D. thank Stefan Willitsch and Michael Drewsen for fruitful discussions. The authors acknowledge support from {}``Agence Nationale de la Recherche'' (ANR), under the project COPOMOL (contract ANR-13-IS04-0004-01).

%\bibliographystyle{unsrt}
%\bibliography{bibliocold_A-K,bibliocold_L-Z,AtoSpec}

%

\end{document}